# Neural Network architectures to classify emotions in Indian Classical Music

Uddalok Sarkar[a]*, Sayan Nag[b], Medha Basu[a], Archi Banerjee[a,c],
Shankha Sanyal[a,d], Ranjan Sengupta[a], Dipak Ghosh[a]

[a]*Sir C.V. Raman Centre for Physics and Music, Jadavpur University, India*
[b]*Department of Medical Biophysics, University of Toronto, Canada*
[c]*Rekhi Centre of Excellence for the Science of Happiness, IIT Kharagpur, India*
[d]*School of Languages and Linguistics, Jadavpur University, India*
*corresponding author, email: uddaloksarkar@gmail.com*

**ABSTRACT**

Music is often considered as the language of emotions. It has long been known to elicit emotions in human being and thus categorizing music based on the type of emotions they induce in human being is a very intriguing topic of research. When the task comes to classify emotions elicited by Indian Classical Music (ICM), it becomes much more challenging because of the inherent ambiguity associated with ICM. The fact that a single musical performance can evoke a variety of emotional response in the audience is implicit to the nature of ICM renditions. With the rapid advancements in the field of Deep Learning, this Music Emotion Recognition (MER) task is becoming more and more relevant and robust, hence can be applied to one of the most challenging test case i.e. classifying emotions elicited from ICM. In this paper we present a new dataset called ***JUMusEmoDB*** which presently has 400 audio clips (30 seconds each) where 200 clips correspond to happy emotions and the remaining 200 clips correspond to sad emotion. The initial annotations and emotional classification of the database has been done based on an emotional rating test (5-point Likert scale) performed by 100 participants. The clips have been taken from different conventional '*raga*' renditions played in *sitar* by an eminent maestro of ICM and digitized in 44.1 kHz frequency. The *ragas*, which are unique to ICM, are described as musical structures capable of inducing different moods or emotions. For supervised classification purposes, we have used 4 existing deep Convolutional Neural Network (CNN) based architectures (resnet18, mobilenet v2.0, squeezenet v1.0 and vgg16) on corresponding music spectrograms of the 2000 sub-clips (where every clip was segmented into 5 sub-clips of about 5 seconds each) which contain both time as well as frequency domain information. The initial results are quite inspiring, and we look forward to setting the baseline values for the dataset using this architecture. This type of CNN based classification algorithm using a rich corpus of Indian Classical Music is unique even in the global perspective and can be replicated in other modalities of music also. This dataset is still under development and we plan to include more data containing other emotional features as well. We plan to make the dataset publicly available soon.
**Keywords: Indian Classical Music, Emotions, Classification, CNN**

# 1. INTRODUCTION

Music imposes emotion. Considering the very basic, major scale always creates a happier ambience whereas minor scale creates a bit sad. This response to melody and rhythm is a biological instinct. A non-musician's ear can feel the pain when Garry Moore plays "Loner", even a baby responds to different scales and melody lines. While talking about emotions elicited by Indian Classical Music (ICM) it always becomes a matter of huge challenge for its ambiguous emotional response. Studies involving non-linear techniques have been conducted in the recent past to understand this complex behavior of music and its manifestation in the human brain [13 - 17].

In the recent years, Machine Learning has made significant advancements in a multitude of fields including computer vision, medical imaging, natural language processing and so on [18 - 37]. Such Machine Learning and Deep Learning approaches have been used to identify different emotions associated with music [9 -12]. So, Music Emotion Recognition (MER) has always been an interesting task to perform for observing the correlation between the music and perceived emotion. Music Emotion Recognition task was first introduced by Barthet et al [5]. Since then, several developments have been achieved on this discipline, and hence it has become very important to observe the role of ICM in elicitation of emotion.

To exploit the significance of emotion induction in ICM, we need a proper database to work with. Previously datasets like Computer Audition Lab 500-song (CAL500) [6], CAL500exp [7], datasets have been introduced, which is enriched with 500 western music clips. Here in this paper we have introduced a database *JUMusEmoDB* enriched with 400 music clips from genre of Indian Classical Music, of which 200 clips fall into the category of happy emotion while 200 falls into sad. Each clip is of 30 seconds length which is long enough for introducing an emotional imposition [8]. All the clips are parts of different '*raga*' renditions improvised in Sitar by an eminent maestro of ICM. Each *Raga* in ICM evokes not a particular emotion (*rasa*), but a superposition of different emotional states such as joy, sadness, anger, disgust, fear etc. To decipher which particular emotion is conveyed in the chosen 30 sec segment of the *raga*, an emotion annotation was performed initially by 100 participants based on 5-point Likert scale.

For an emotion classification task different acoustic features of music are very important. Different acoustic feature consists of (a) Rhythmic Features: Tempo, Silence etc.; (b) Timbral Features: MFCC, Average Energy, Spectral Centroid, Spectral Tilt etc.; (c) Chroma Features. These acoustic features quantify the musicality of a clip which in effect contributes to MER. Study on this knowledge driven approaches results in an efficient model structure; but before this, a validated dataset is very necessary to work with. After the emotion annotation, a data driven approach has been used to validate our dataset.

We have taken an image processing-oriented approach to classify the dataset into emotion tags. We primarily extracted the spectrogram of a clip and fed the processed spectrogram image into existing deep Convolutional Neural Network (CNN) based architectures. These CNNs were then trained to classify emotions. For this study, we have made use of four different existing CNN architectures: VGG16, ResNet18, MobileNet v2.0, SqueezeNet v1.0 and have received some promising results. Furthermore, the dataset used in our study which we have named as *JUMusEmoDB*, is a novel dataset comprising of clips from Indian Classical Music genre. This dataset currently has musical clips from two emotions, namely happy and sad, and is still under development. We plan to include more data containing other emotional features eventually making the dataset publicly available shortly. This dataset can be used in future by the scientific community for emotion classification purposes, to investigate the impact of Indian Classical Music on human brain and finally to conduct cross-cultural studies combining both Western Classical and Indian Classical musical clips.

The paper has been organized as follows: Section 2 contains Data Acquisition, Section 3 comprises of the Methods used, the Experiments and Results of the study have been mentioned in Section 4 and finally Section 5 has the conclusion.

## 2. DATA ACQUISITION:

*JUMusEmoDB* consists of 400 audio clips of 30 second each. The clips have been taken from different conventional *raga* renditions played in *sitar* by an eminent maestro of ICM and recorded with a sample frequency of 44.1 kHz. 200 clips correspond to happy emotions and the remaining 200 clips correspond to sad emotion. Initial annotations and emotional classification of the database has been done based on an emotional rating test (5-point Likert scale) performed by 100 participants.

## 3. METHODS:

We have labeled our music database *JUMusEmoDB* into two main classes as stated earlier, i.e., happy, and sad. In this paper we have followed the approach of a data driven method to classify the database into two distinct classes. Now a general question arises regarding why a data driven approach is being followed primarily rather than any novel knowledge-based quantification. To answer this, we have to take into account that a successful classification with high accuracy rate will validate the authenticity of *JUMusEmoDB*, which can then be used to develop new knowledge-based models to test on this database.

### 3.1. Data Preprocessing

The basic input to our CNN based framework should be an image data. Hence, to map our audio database into image paradigm we have made use of Spectrogram. But before obtaining spectrogram for these 30-second-long audio clips, we have sliced all the 30 second clips into individual 5 second clips to augment our dataset and performed STFT on this augmented dataset to obtain spectrograms.

### 3.2. Spectrogram

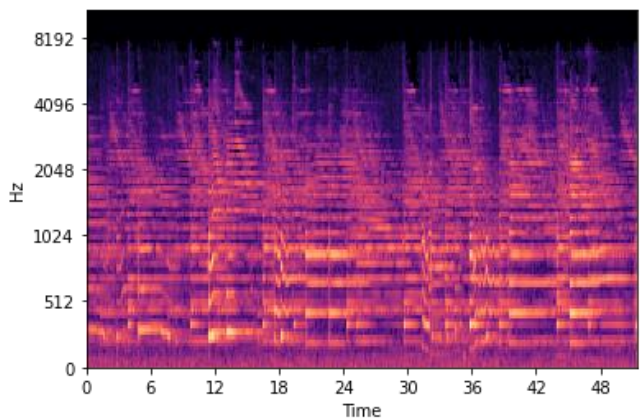

*Fig.1: Spectrogram of a HAPPY Music Clipping*

In our proposed framework, we have performed our classification task on the mel-spectrograms of derived music tracks. To extract the melspectrogram we have made use of short time Fourier transform with a window size of 2048 and hop size of 512 to obtain a spectrogram (**Fig. 1**). Thus we obtained melspectrogram as a dot product of obtained spectrogram with mel filterbanks.

$$spect(t,f) = |stft(t,f)|^2$$

$$melspectrogram(t,f') = spect(t,f) \cdot melfilterbanks$$

## 3.3. CNN Models

We have extended our framework into four different established ConvNet models, i.e., VGG[1], ResNet[2], SqueezeNet[3], MobileNet v2[4]. We thus have obtained four accuracy rates with an average of 99.117%. We have added an extra layer of 2 channels at each end of considered model because of bi-class output of our framework.

**3.3.1 VGG Net** is the oldest architecture among the used ConvNet models, proposed by Karen Simonyan and Andrew Zisserman of Oxford Robotics Institute in 2014 [1]. This has sixteen weight layers; thirteen convolution layers divided into five groups, each group followed by pooling layers, and three fully connected (FC) layers at the end of whole network (**Fig. 2**). Convolution layers have a receptive field of 3x3 throughout the whole net, with stride 1. The Maxpooling layers consist of receptive fields of size 2x2 each and with a stride of 2. The network ends with three fully connected layers with first two layers of 4096 channels and last layer of 1000 channels due to 1000 classes of ILSVRC (ImageNet Large-Scale Visual Recognition Challenge).

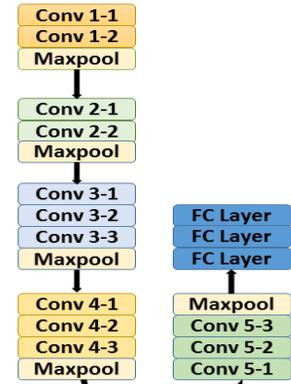

*Fig.2: VGG-16 Network Architecture*

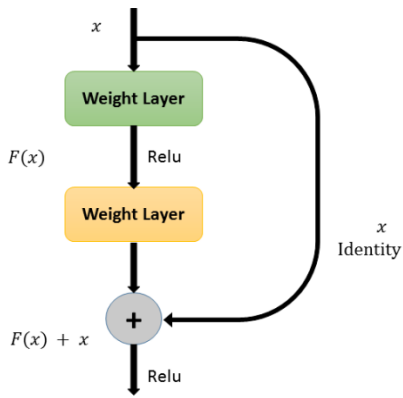

*Fig.3: Residual Block*

**3.3.2. ResNets** are residual learning framework with substantially deeper network but with lower complexity. Kaiming He, Xiangyu Zhang, Shaoqing Ren, Jian Sun has proposed a residual network with a depth of up to 152 layers (**Fig. 3-4**) i.e., 8x deeper than VGG-19 nets [2]. Increasing depth of a network can lead to a very serious problem of vanishing gradient which results in saturation of convergence with a very high training error and low accuracy problems. Kaiming He et al. has beautifully taken care of these facts and modified a very deep network to gain a high accuracy with low training error. They have implemented a "short-cut connection" of identity mapping. Their approach was to allow the network to fit the stacked layers to a residual mapping using residual block (**Fig. 3**) instead of fitting them directly to the underlying mapping. So, instead of feeding $H(x)$ (desired underlying mapping) let the stacked network fit, $F(x) := H(x) - x$ and thus ultimately gives $H(x) := F(x) + x$.

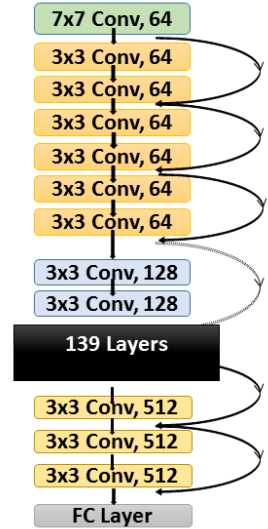

*Fig.4: ResNet 152-layer Architecture*

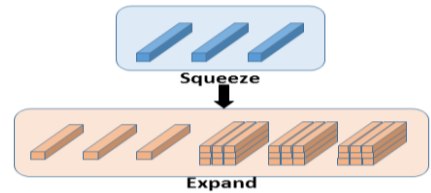

*Fig.5: Fire Module*

**3.3.3. SqueezeNet** is a lighter modification of deep convolutional neural networks which has achieved an accuracy near to AlexNet (on ImageNet dataset) with 50 times fewer parameters (**Fig. 6**) [3]. The main concept behind this architecture is introduction of 'fire module". A fire module is a stacking of a squeeze layer with 1x1 convolution filters and an expand layer which has both 1x1 and 3x3 filters. Number of kernels in squeeze layer should be less than number of kernels in expand layer to limit the number of input channels to 3x3 kernels. **Fig. 5** shows an architectural view of fire module.

**3.3.4. MobileNet,** by Google, has introduced a new kind of lightweight architecture by replacing traditional convolution layer with "Depth-wise Separable Convolution" to reduce the model size and complexity. In MobileNetV2 two kinds of blocks are observed [4], stride 1 block (residual block), stride2 block (downsizing). Each block consists of three layers as shown in **Fig. 7**. First layer contains 1x1 kernel with RELU6. Second layer performs the depth-wise convolution. Third layer again contains 1x1 kernel without any nonlinear function.

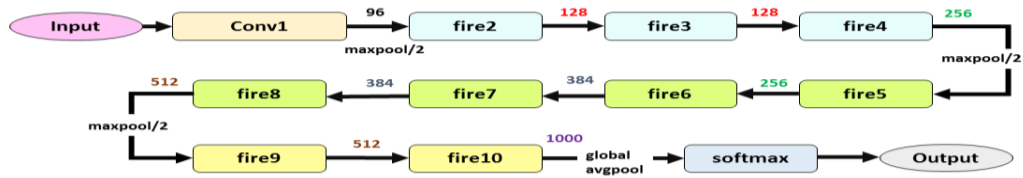

*Fig.6: SqueezeNet Architecture*

## 3.4. Emotion Extraction from Network output

Output from the fully connected layers is provided to softmax to extract out the classes. In our framework, we are concerned about two distinct classes (i.e., Happy Music clip, Sad Music clip). All the four architectures contain 1000 channels in last fully connected networks by default 1000 classes of ILSVRC (ImageNet Large-Scale Visual Recognition Challenge). So, we added another fully connected layer of 2 channels at the end to cater our purpose. After this last fully connected layer a softmax is performed to extract out the emotion classes.

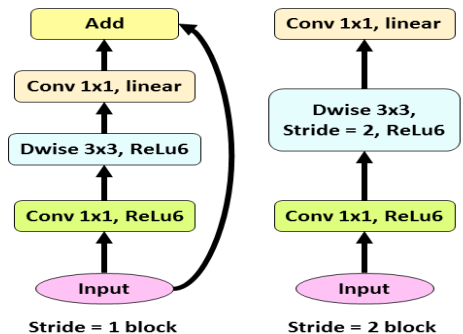

*Fig.7: MobileNet Architecture*

## 4. EXPERIMENTS AND RESULTS

As a classification task, some criteria are being used to quantify the classification performance. MSE (mean squared error), MAE (mean absolute error) has been used a lot as loss functions for image purposes. In our classification task we have made use of cross entropy loss. Entropy is a measure of uncertainty, and it is measured as,

$$H(X) = \begin{cases} -\int p(x)\log(p(x)), & x \text{ is continuos random variable} \\ -\sum p(x)\log(p(x)), & x \text{ is discrete random variable} \end{cases}$$

Cross entropy loss (by using the idea of Entropy) measures the resemblance of actual output against predicted output. Cross entropy increases along with the divergence of prediction from actual output. Hence a 0 loss represents perfect model. Cross entropy loss or Log loss is being calculated using the following equation:

$$L = -\sum_i y_i \log(\hat{y}_i) w$$

where $y_i$ is the calculated or predicted output and $\hat{y}_i$ is the actual output.

Previously we have introduced our used framework and model improvisation. In this section we are illustrating the acquired results. For our work, we used a train/validation split ratio as 85/15. In the training phase we receive a model cost (calculated from loss function) which indicates the model performance. The loss curve for training phase for each model is shown in the Fig. 8. With time (iterations) convergence is achieved for each of the models used and the performance of SquuezeNetV1 in the training phase also outperforms the other three models used.

The convergence plot helps to tune the parameters of CNN models. After adjusting the parameters, we obtain the best results for each CNN Models. **Table 1** shows the validation accuracy of the aforementioned models.

| Model | Accuracy |
|---|---|
| VGG16 | 99.007% |
| ResNet18 | 97.682% |
| SqueezeNetV1 | **99.669%** |
| MobileNetV2 | 98.675% |

*Table 1: Accuracies of different CNN models on the validation dataset*

From Table 1, we can see that SqueezeNetV1 seems to fit best with the acquired dataset. The dataset size (as of now) is not that huge which makes it suitable for light weight models (with fewer parameters) like SqueezeNetV1 which gives the best validation accuracy.

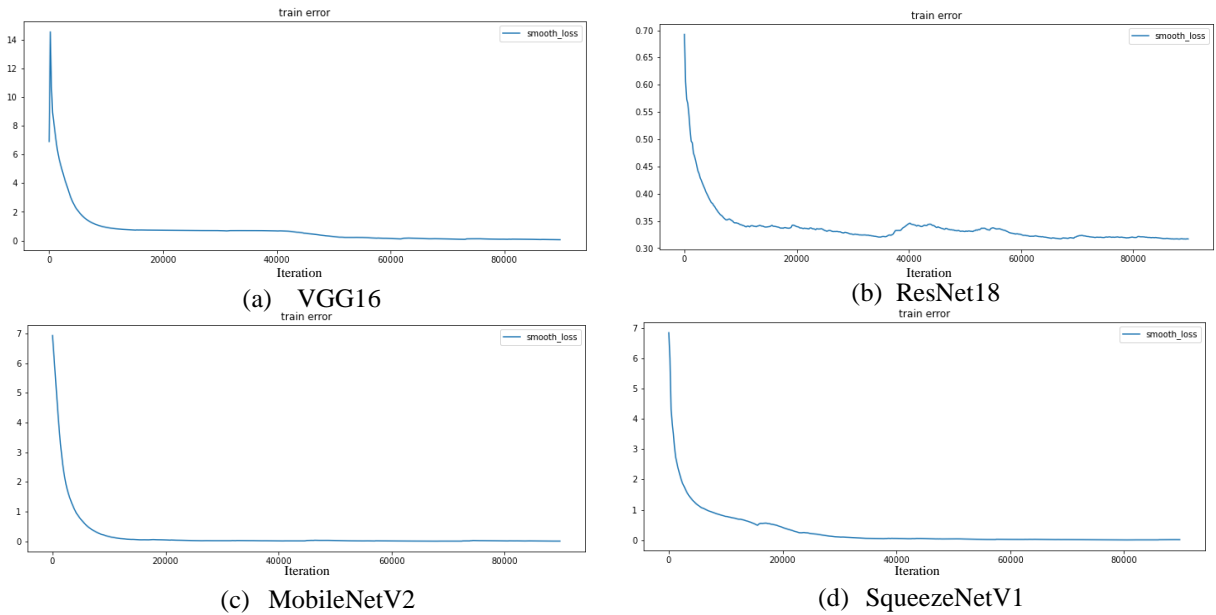

(a) VGG16
(b) ResNet18
(c) MobileNetV2
(d) SqueezeNetV1

*Fig.8: Training Loss curves*

## 5. CONCLUSION

In this work, we proposed a novel dataset called *JUMusEmoDB* which presently has 400 audio clips (30 seconds each) where 200 clips correspond to happy emotions and the remaining 200 clips correspond to sad emotion. The initial annotations and emotional classification of the database has been done based on an emotional rating test (5-point Likert scale) performed by 100 participants. We also demonstrated the performances of four deep CNN based architectures namely resnet18, mobilenet v2.0, squeezenet v1.0 and vgg16. Validation accuracy values showed that SqueezeNetV1 performed the best out of the four models. Even though the advantage of employing CNN based architectures for tackling the problem of music emotion recognition include better overall accuracy because of better extraction of useful features from the data compared to other traditional methods, further studies need to be conducted to understand the source of emotions for a given music. Another limitation of this work is the lack of data in our dataset, but this is a pilot study it is still under development. We plan to incorporate more data containing other emotional features as well and eventually make the dataset publicly available shortly.

## ACKNOWLEDGEMENT

Archi Banerjee acknowledges the Department of Science and Technology (DST), Govt. of India for providing the DST CSRI Post Doctoral Fellowship (DST/CSRI/PDF-34/2018) to pursue this research work. Shankha Sanyal acknowledges DST CSRI, Govt of India for providing the funds related to this Major Research Project (DST/CSRI/2018/78 (G)) and the Acoustical Society of America (ASA) for providing the International Students Grant.